\documentclass[twocolumn,tighten,times,floatfix]{aastex631}
\usepackage{xspace,xcolor}
\usepackage{natbib}
\usepackage{outlines}

\usepackage[textwidth=1.4cm, textsize=small, colorinlistoftodos]{todonotes}

\usepackage{amsmath}
\usepackage{graphicx}
\usepackage[utf8]{inputenc}

\usepackage{enumerate}

\usepackage{hyperref}
\usepackage[capitalize, noabbrev]{cleveref}

\pdfoutput=1

\crefname{section}{\S}{\S\S}

\makeatletter
\usepackage{etoolbox}
\patchcmd\H@refstepcounter{\protected@edef}{\protected@xdef}{}{}
\makeatother

\newcommand{\beq}{\begin{equation}}
\newcommand{\eeq}{\end{equation}}

\newcommand{\sne}{SNe~Ia }

\newcommand{\JHU}{Department of Physics and Astronomy, Johns Hopkins University, Baltimore, MD 21218, USA}
\newcommand{\STScI}{Space Telescope Science Institute, Baltimore, MD 21218, USA}
\newcommand{\Morgan}{Morgan State University, Baltimore, MD 21251, USA}
\newcommand{\ISEF}{ISEF International Fellowship}

\revised{July 2024}
\accepted{for publication in The Astrophysical Journal}

\begin{document}

\title{The Reliability of Type Ia Supernovae Delay Time Distributions Recovered from Galaxy Star Formation Histories}

\shorttitle{SN~Ia DTDs from IllustrisTNG}
\shortauthors{Joshi, Strolger, \& Zenati}

\author[0000-0002-7593-8584]{Bhavin A.~Joshi}
\affiliation{\JHU}

\author[0000-0002-7756-4440]{Louis-Gregory Strolger}
\affiliation{\STScI}
\affiliation{\JHU}
\affiliation{\Morgan}

\author[0000-0002-0632-8897]{Yossef Zenati}
\altaffiliation{\ISEF}
\affiliation{\JHU}
\affiliation{\STScI}

\correspondingauthor{Bhavin A.~Joshi}
\email{bjoshi5@jhu.edu}

\begin{abstract}
We present a numerical analysis investigating the reliability of type Ia supernova (SN~Ia) delay-time distributions recovered from individual host galaxy star-formation histories. We utilize star-formation histories of mock samples of galaxies generated from the IllustrisTNG simulation at two redshifts to recover delay-time distributions. The delay-time distributions are constructed through piecewise constants as opposed to typically employed parametric forms such as power-laws or Gaussian or skew/log-normal functions. The SN~Ia delay-time distributions are recovered through a Markov Chain Monte Carlo exploration of the likelihood space by comparing the expected SN Ia rate within each mock galaxy to the observed rate.
We show that a reduced representative sample of \emph{non-host} galaxies is sufficient to reliably recover delay-time distributions while simultaneously reducing the computational load.
We also highlight a potential systematic between recovered delay-time distributions and the mass-weighted ages of the underlying host galaxy stellar population.
\end{abstract}

\section{Introduction}
\label{sec:intro}
Type Ia supernovae (SNe Ia) are essential cosmological probes of the expansion rate of the universe and the nature of dark energy, yet there remains some ambiguity on what systems give rise to such important events \citep[see,][]{Phillips93, Riess+98, Perlmutter+99}. Indeed, the exact mechanism and progenitor systems of SNe Ia remain ongoing topics of debate after decades of investigation. Broadly, the main scenarios are the single degenerate (SD) model, a white dwarf (WD) accreting material from a companion star until it reaches the Chandrasekhar mass limit $M_{ch} \sim 1.4 M_\odot$, causing a thermonuclear explosion, and the double degenerate (DD) model, where two Carbon-Oxygen WD (CO WD) merge due to loss of orbital angular momentum and energy through the emission of  gravitational--waves. There also remains some ambiguity on the robustness of these mechanisms to chemical enrichment history, and therefore historical variances \citep{Hoyle&Fowler60, Arnett69, Whelan&Iben73, Phillips93, LivioM00, Timmes2003, Mazzali+07, Maoz14, Livio&Mazzali18, Ropke+19_book}. No one is sure how these play out within the variety of event characteristics often seen, thereby offering little empirical culling to isolate pristine samples. 

Several models for \sne have been studied and discussed, but none have been able to consistently reproduce the observed properties of SNe, nor their host galaxy rates and distributions \citep{Ruiter+09, Foley+10}. It has been suggested that \sne may come from multiple progenitor channels with wide agreement on the critical mass limit of the WD, known as the Chandrasekhar limit (although, \sne could also happen in a sub-Chandrasekhar scenario) \citep[e.g.,][]{Nomoto+84, Hoelflich&Khokhlov96, Foley+10, shen2010, MariusDan+11, Maoz14, PeretsZenati+19, PolinA+19, Ruiter20}. This includes a single degenerate (SD) model \citep[e.g.,][]{Whelan&Iben73}, a double degenerate (DD) model \citep[e.g.,][]{IbenTutukov84, li1997evolution, Guillochon+10}, along with several other theoretical scenarios to produce a type Ia supernova explosion \citep[e.g.,][]{Hoeflich+95, Hillebrandt&Niemeyer00, Raskin+09, KashiSoker11, Pakmor+11, Pakmor+13, ShenK+18, Soker19, Jha+19, Tanikawa+19, ZenatiFisher20, Pamkor+22_DD, Casabona&Fisher24}.

Supernovae delay-time distributions (DTDs) provide an insight into these progenitor problems, as they provide the distribution of time scales necessary to go from an instantaneous burst of star formation to explosion \citep[see][]{Madau98, Yungelson&Livio00, Han&Podsiadlowski04, Mennekens10, Strolger20, Wiseman21}. These timescales are convenient as they capture several phenomenological epochs in the lifetime of a potential progenitor system, including main-sequence lifetimes, mass accretion-rate timescales to reach Chandrasekhar mass, angular momentum-loss timescales, and many others, which also are dependent on initial conditions, e.g., initial mass function, binarity fractions, and distributions of initial separations.
In order to identify the DTDs of the two main scenarios we need to take into account their evolution paths. 

Attempts to recover these important distributions have typically relied on bulk analyses of supernova rates in comparisons to host properties such as current rate of star-formation or total mass \citep[e.g.,][]{Mannucci05, Mannucci06, Scannapieco05, Ruiter+09}. While these mass-weighted, luminosity-weighed, and star-formation-rate-weighted ``SNu" {\footnote{The SN frequency unit SNu, which is defined as 1 SNu = 1 SN per century per $10^{10} \, \mathrm{L}_\odot$ in the B--band, and measures like it have been in use for some time to measure SN rates \citep[e.g.,][] {Tammann1970, Tammann1974, Tammann1982}.}}
rates have proven to be useful by way of establishing the likelihood of both a prompt and delayed component to SN~Ia progenitors, these methods lack a direct tie to the detailed binary synthesis modeling as they average over a host population, or oversimplify complex star-formation histories (SFHs), and therefore lose a clearly established connection to SN~Ia progenitor systems. Similarly, the ``A+B'' model \citep{Scannapieco05} for considering SN~Ia rates oversimplifies the connection of SNe~Ia to their progenitor environments. More useful methods have compared SN rates directly to star formation rate histories in volumetric studies~\citep{Maoz11}. But those too lose the complexity of star-formation histories, in favor of a more generalized cosmic star formation density history, and likely similarly dilute any subtleties that hint at more than one progenitor mechanism for SN Ia populations. 

As has been demonstrated by many previous studies, one promising method for using all of the information in the SFHs of individual galaxies was introduced in \citet{Maoz11} (with a similar method presented a little earlier in \citealt{Brandt2010}), which presented the recovery of the delay time distribution as a linear, maximum-likelihood recovery problem. Through this maximum-likelihood method one could use SFHs, however detailed by the modeling of galaxies used, to recover DTDs of similar detail, limited by (a) the quality of the measured host galaxy SED, (b) the level of detail used in recovering the SFHs in either parametric or non-parametric methods, and (c) the level of detail in the modeling of DTDs, whether parametric or non-parametric. As discussed more in later sections, the method recovers the best-fitting DTD model by maximizing the likelihood (actually the log-likelihood) that galaxies that hosted SNe would do so, based on a present peak in their \sne rate histories, and that these heightened likelihoods stand out from the noise of the likelihoods from \emph{non-host} galaxies.

Here we present a thorough analysis of model galaxies as a precursor to a full-up analysis of \texttt{Pantheon} + SN~Ia host galaxies \citep{Scolnic22, Brout22}. We analyze the inferred SN~Ia DTD from various simple host selection criteria to: (a) understand potential systematics by employing a sample with well-understood properties, and (b) improve the methods' computational performance.
A potential systematic effect that we observe is likely caused by the timescales at which the ``parent population'' of galaxies that SN~Ia hosts are drawn from, made most of their stars. We also present and discuss systematic effects that can arise when fitting for a potentially large spread of delay-time distributions with parametric models, such as a power law, constructed from a small number of fixed parameters.
These systematics can potentially affect the calibration of uncertainty budgets for current and future studies that use SN~Ia as a probe of dark energy.

Previous studies in the literature have employed relatively simple forms for the DTD with a power law being most common ($t^{-\alpha}$; \citealt{Totani08,  Graur11, Graur14, Rodney14, Friedmann18, Heringer19, Freundlich21, Wiseman21}), along with Gaussian and skew-normal functions as well \citep[e.g.,][]{Strolger04, Strolger20}.
Many of these studies have provided evidence for a prompt component of delay-times producing type Ia supernovae shortly after star formation with a large contribution at times $\lesssim$1 Gyr. While a large prompt contribution is expected from simulations of the evolution of white dwarf binary systems \citep[e.g.,][]{Ruiter+09, Mennekens10}, and the Fe content of cluster galaxies \citep[e.g.,][]{Maoz10}, a ``tail'' of longer delay times ($>$2 Gyr) is also well supported by DTD reconstruction methods.

In this work, we favor ``non-parametric'' DTD models (analogous to non-parametric SFHs) in lieu of simpler parametric forms, because a non-parametric binned DTD is not restricted by the form of the mathematical expression. The remainder of the paper is organized as follows: we provide details of the IllustrisTNG simulation and the sample we choose from it in \cref{sec:sample}, in \cref{sec:like} we discuss the likelihood-based method and its implementation, in \cref{sec:tests} we outline our test cases and the host selection criteria, we present our results in \cref{sec:results} and a discussion of systematic effects observed in the recovered DTDs in \cref{sec:discussion}, and we conclude with a summary of this work in \cref{sec:summary}. Where required, we have assumed a flat $\Lambda$CDM Universe with $\Omega_M=0.3$, $\Omega_\Lambda=0.7$, and $H_0=70$ km/s/Mpc.

\begin{figure*}
    \centering
    \includegraphics[width=0.9\textwidth]{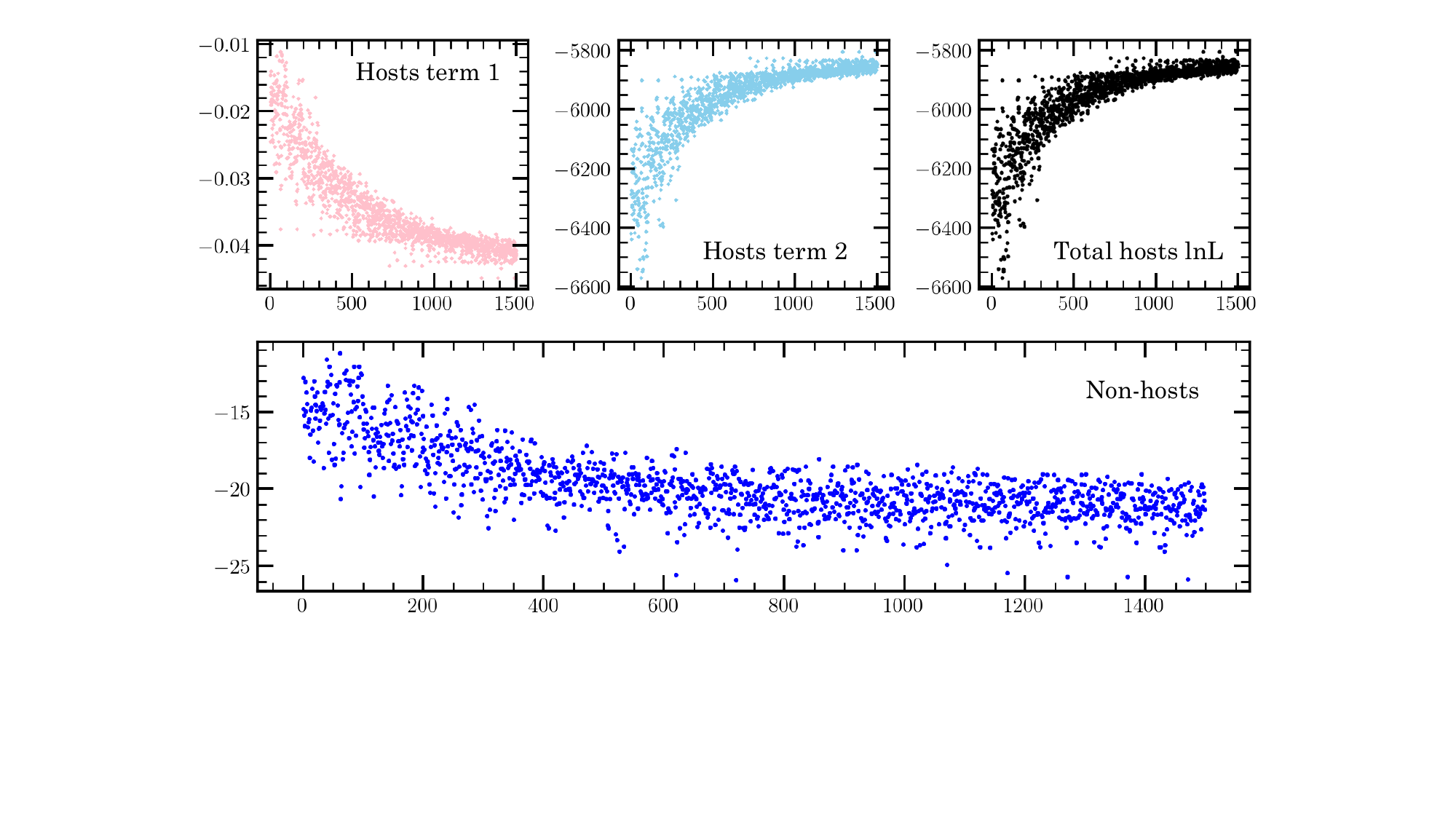}
    \caption{Individual terms of the DTD log--likelihood computation (eqn.~2 and 3). The x-axis for each panel shows the step number for the MCMC walkers. The terms for host galaxies are: term 1 = $-\sum^{N_\mathrm{hosts}} m_i$ and term 2 = $\sum^{N_\mathrm{hosts}} \mathrm{ln} \left( m_i^{n_i}/n_i!\right)$. The sum of these two terms gives the total host log-likelihood. The \emph{non-host} galaxy log-likelihood is given by: $-\sum^{N_\mathrm{non\text{-}hosts}} m_i$ (eqn.~3).}
    \label{fig:lnL}
\end{figure*}

\section{Illustris-TNG sample}
\label{sec:sample}
The IllustrisTNG (hereafter TNG) simulations are a suite of large-scale, cosmological magnetohydrodynamical simulations of galaxy formation and evolution \citep{Weinberger17, Pillepich18a, Nelson19}. Of the three simulated volumes of 50, 100, and 300 Mpc$^3$ we employ the TNG300 box because it has the largest number of galaxies/subhalos available for analysis within the simulated volume. Each TNG simulation contains baryonic and dark components (along with dark-only counterparts) and we utilize the highest mass resolution baryonic + dark TNG300-1 simulation. We choose to work with the TNG simulations given the large simulated volume and their inclusion of state-of-the-art descriptions of galaxy formation and evolution physics. More importantly, TNG reproduces observed phenomena and scaling relations
-- e.g., (i) galaxy color bimodality \citep{Nelson19}, (ii) matter and galaxy clustering \citep{Springel18}, (iii) stellar mass content of groups and clusters of galaxies \citep{Pillepich18b}, (iv) chemical evolution of galaxies \citep{Naiman18}, and (v) galaxy cluster radio halos and magnetic fields scaling relations \citep{Marinacci18}.

The total number of galaxies in our sample is $2.5\times 10^5$ at each of the two redshifts we chose to work with, $z=0$ and $z=1$. We arrive at the chosen sample of galaxies by applying the following conditions to the TNG300 snapshot data at each redshift:
(i) minimum number of star particles set at 50, and
(ii) total subhalo mass within the mass range $10^8 \leq \mathrm{M_s\,[M_\odot]} \leq 10^{12}$.
It should be noted that while more than $2.5\times 10^5$ galaxies pass our selection criteria at each redshift, our sample selection code stops at this number, due to computational constraints, having reached a statistically significant sample size. The SFH for any galaxy is constructed by summing over the initial stellar masses formed in each lookback time bin for each individual star particle associated with the galaxy. 

\section{Methods}

\subsection{Likelihood-function based DTD recovery}
\label{sec:like}

The DTD recovery method we consider here was introduced by \citet{Brandt2010} and \citet{Maoz11}, as a prescription that into account the SFHs of \emph{all} individual galaxies within a survey (including \emph{non-host} galaxies). The DTDs are recovered by employing all available information within a survey (or a combination of surveys) in a statistically rigorous manner \citep[e.g.,][]{Maoz12, Graur13}. This method, with a straightforward log-likelihood function, is particularly well-suited for a maximum likelihood estimation through a Markov Chain Monte Carlo (MCMC) exploration of the DTD parameter space. 
Following the prescription detailed by 
\citet{Strolger20}, the probability of observing $n_i$ SN~Ia in the $i^{th}$ galaxy given that the expected number is $m_i$ follows a Poisson distribution, $P(n_i|m_i) = (m_i^{n_i}e^{-m_i})/n_i!$. Therefore, for `$N$' galaxies in a survey the log-likelihood is given by,
\begin{equation}
    \mathcal{L} = \Pi^N_i\, P(n_i|m_i) \implies \mathrm{ln} \mathcal{L} = -\sum^N m_i \, + \, \sum^N \mathrm{ln} \left(\frac{m_i^{n_i}}{n_i!}\right).
\end{equation}
The expected number of SN~Ia in any galaxy ($m_i$) is derived from the convolution of the SFH of the galaxy and the \emph{global DTD}, along with multiplicative factors specifying the fraction of the initial mass function that contributes to stellar masses between 3-8$M_\odot$ and the fraction of these that can be expected to end as SN~Ia (for details see eqn.~8 in \citealt{Strolger20}).

The above log-likelihood expression is a sum of host and \emph{non-host} terms as follows:

\begin{align}
    \mathrm{ln} \mathcal{L} (\mathrm{hosts}) &= -\sum^{N_\mathrm{hosts}} m_i \, + \, \sum^{N_\mathrm{hosts}} \mathrm{ln} \left(\frac{m_i^{n_i}}{n_i!}\right); \ n_i \geq 1 ;\\
    \mathrm{ln} \mathcal{L} (\mathrm{non\text{-}hosts}) &= - \sum^{N_\mathrm{non\text{-}hosts}} m_i; \ n_i = 0
\end{align}

with $\mathrm{ln} \mathcal{L} = \mathrm{ln} \mathcal{L} (\mathrm{hosts}) + \mathrm{ln} \mathcal{L} (\mathrm{non\text{-}hosts})$.
Since most galaxies in a survey will not host a SN~Ia, it is possible to significantly reduce the computational expense of recovering DTDs by avoiding inferring SFHs for \emph{non-host} galaxies and then subsequently employing these SFHs in the \emph{non-host} log-likelihood term shown above. More details are provided in the Appendix where we show that we can reliably recover the DTD even with a reduced sample of \emph{non-host} galaxies.

Rather than assume a parametric form for the DTD (power-law, exponential, etc.), for this work we employ a non-parametric eight-bin DTD model with equal weight given to each bin, i.e., $\sum_i \phi_i = 1$. The time bins we chose are [in Gyr]: 0--0.5, 0.5--1, 1--2, 2--4, 4--6, 6--8, 8--10, and 10--13.7.
These time bins provide the DTD inference method adequate leverage to distinguish between prompt, delayed, or intermediate delay times. We chose these eight-time bins after some trial and error to determine the number of bins and range for each bin. This DTD form gives us enough flexibility to distinguish between different progenitor scenarios while still being computationally efficient. An added advantage is that any secondary or tertiary populations of delay--times (that are not extremely prompt or delayed) can be revealed by the non-parameterized DTD forms, therefore providing us valuable clues to SN~Ia progenitors.

We show the progression of the separate log-likelihood (lnL) terms in \cref{fig:lnL} for one of our test cases (see \cref{sec:tests}), where host galaxies are chosen to exhibit star formation at very early times ($>$10 Gyr). The values shown belong to different terms in equations (2) and (3) as the MCMC walkers explore the DTD parameter space. Convergence towards the final recovered DTD is quite clear and relatively rapid. It is also clear that the dominant term driving the lnL evolution is the $\sum^N\mathrm{ln}(m_i^{n_i}/n_i!)$ term for the host galaxies. A similar behavior is observed for the individual lnL terms for other test cases as well. We build on this idea and show that a reduced number of \emph{non-host} galaxies can still be used to reliably infer the DTD for each test case (see Appendix).

\begin{figure*}
    \begin{center}
    \includegraphics[width=\textwidth]{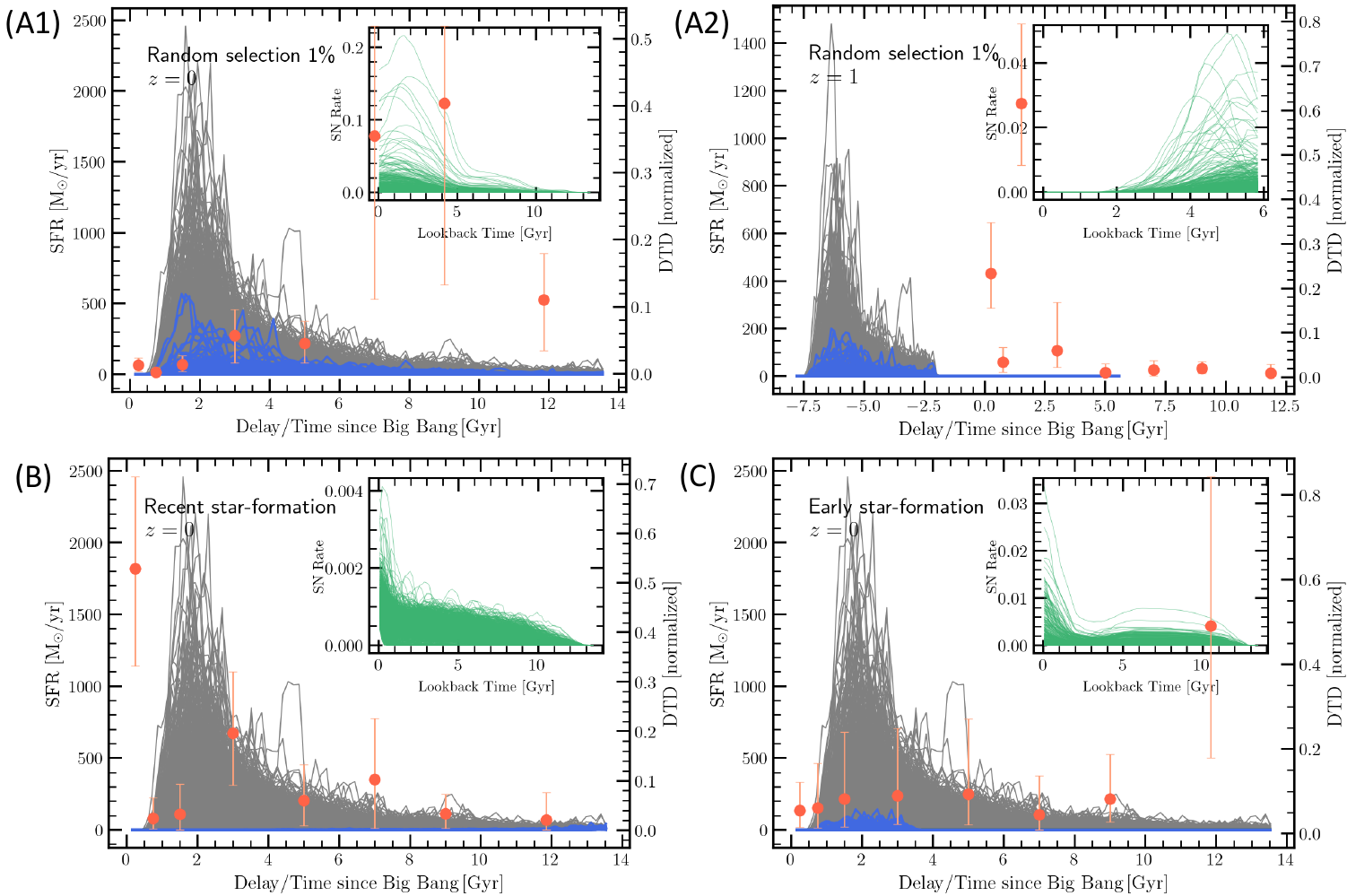}
    \caption{DTDs recovered for the full sample size of 250000 galaxies with host selection criteria as shown. The DTDs are shown as orange points and the SFHs for the \emph{non-host} and host galaxies are shown in gray and blue, respectively. The panel number corresponds to the case number shown in \S\ref{sec:tests}. The inset plots in each panel show the SN rate history for host galaxies.}
    \end{center}
    \label{fig:dtd_rec_fullsample}
\end{figure*}

\subsection{Diagnostic DTD cases} 
\label{sec:tests}
To explore the full shape of the DTD for SN~Ia we employ simulated TNG galaxies for the cases given below. 
In each case, we assign the selected host galaxies to host a SN~Ia at the observed epoch.
\begin{itemize}
    \item { Case A1: Host galaxies chosen at random from full TNG sample \emph{at low-$z$}.} We choose a host fraction of 1\% with host galaxies chosen at random. The random selection implies host galaxies should be representative of the full sample. The results are shown in \cref{fig:dtd_rec_fullsample} panel (A1).

    \item { Case A2: Host galaxies chosen at random from full TNG sample \emph{at high-$z$}.} Same as case A1 but galaxies chosen from a higher redshift TNG snapshot ($z=1$). The results are shown in \cref{fig:dtd_rec_fullsample} panel (A2).

    \item { Case B: Short delay, $\lesssim$ 2 Gyr.} We now choose host galaxies that have significant star-formation at lookback times less than 2 Gyr ensuring that delay times peak at times less than 2 Gyr. The results are shown in \cref{fig:dtd_rec_fullsample} panel (B).

    \item { Case C: Long delay, $\geq$ 10 Gyr.} In this case, we choose host galaxies that have exactly zero star formation at lookback times less than 10 Gyr ensuring that delay times are longer than 10 Gyr. The results are shown in \cref{fig:dtd_rec_fullsample} panel (C).

\end{itemize}

\section{Results}
\label{sec:results}

\begin{figure}
    \centering
    \includegraphics[width=0.45\textwidth]{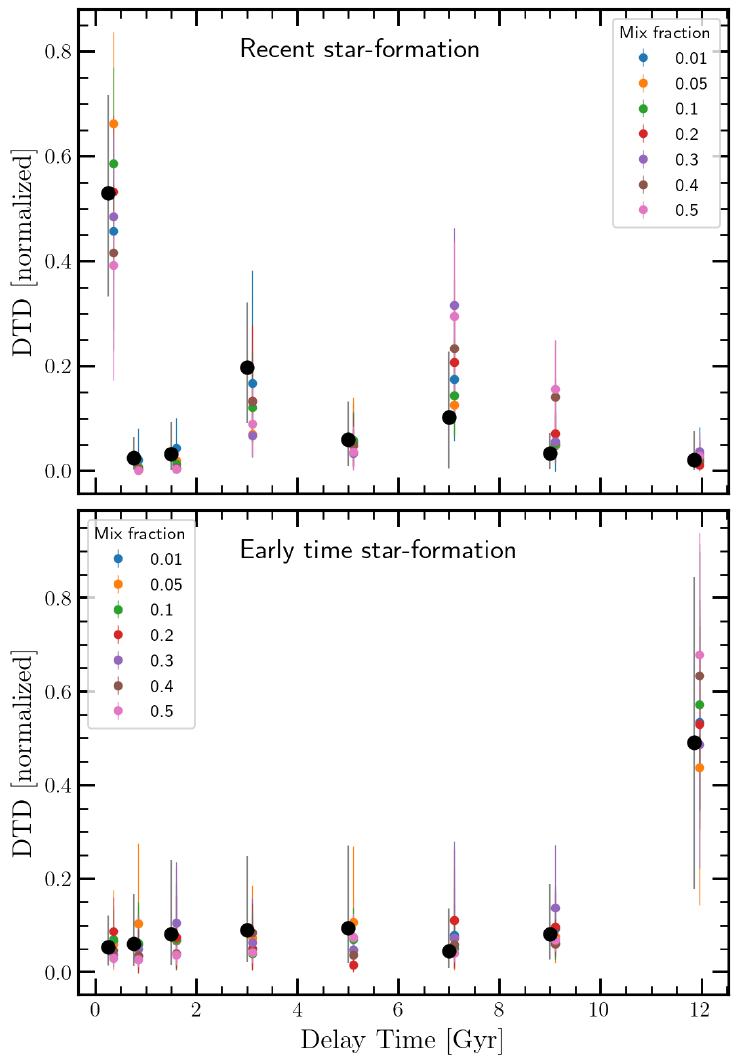}
    \caption{The recovery of SN~Ia DTDs from cases B and C when host galaxies (satisfying selection from case B and C) are mixed with additional hosts from background galaxies that do not satisfy the selection criteria. The top and bottom panels showcase B and C corresponding to recent and early-time star formation, respectively. The fraction of background galaxies added to hosts is shown in the legend with the reference case (without any mixing) shown with black points.}
    \label{fig:mix_recovery}
\end{figure}

We recover the DTD in each test case by using the full sample of 250K galaxies as shown in \cref{fig:dtd_rec_fullsample}. In the Appendix we also show DTD recovery from a reduced number of \emph{non-host} galaxies without any loss of reliability (\cref{fig:dtd_rec_reducedsample}, with 50K non host galaxies) but with a dramatic reduction in computational expense. The results for each case for the full sample run are tabulated in \cref{tab:summary}. The table also shows the best fit power-law index (along with a multiplicative coefficient) for each case.

Additionally, in order to test the reliability of recovered DTDs in the presence of host galaxies which have been wrongly assigned to be host galaxies we run tests by adding galaxies that were previously \emph{non-hosts} to the originally selected host galaxies. We show these DTD recovery results from the addition of confounding
background galaxies to the hosts selected specifically for late and early time star-formation (cases B and C above) in \cref{fig:mix_recovery}. The ``mix'' fraction of confounding galaxies ranges from 1\% to 50\% of the original host galaxies. In each run it can be seen that the DTD recovered is consistent with the original DTD (i.e., DTD inferred without any added background galaxies).

Cases A1 ($z=0$) and A2 ($z=1$) are chosen to be more suitable for SN~Ia detected in galaxy redshift surveys where the host galaxies are ``representative'' of the background population of galaxies at that redshift. This situation is not an exact reproduction of physical reality, because SN~Ia host galaxies are unlikely to be perfectly representative of the background population of non host galaxies. Rather, there is some expectation that they should be more ``active'' than most background galaxies, with high SN rate likelihoods. However, in a case where a sample of host and non host galaxies comes from a survey, if some fraction of the host galaxies are representative of the background population of galaxies this could indicate that some non-zero weight should be assigned to the DTD resulting from the background population.
Another thing to note is how different the DTDs for cases A1 and A2 are from each other. For a host galaxy sample that spans a wide range in redshift this could imply an underlying DTD that is dependent on redshift. Attempting to infer a \emph{global DTD} from such a sample will likely entangle very different DTDs causing confusion for the resulting DTD.

For case B, while the DTD recovery for the primary host galaxy selection is robust, i.e., the DTD bins where highest weights are recovered are expected to have the highest weights given the host selection (\cref{fig:mix_recovery}), there also appears to be a significant secondary DTD component in the form of a broad distribution of longer delay times peaking around $\sim$5 Gyr. Another approach with the recent star-forming hosts, is the substantial ``tail'' of longer delay times in the DTD appears consistently in all our runs (with and without mixing; \cref{fig:mix_recovery}).
While we avoid a physical interpretation of the secondary DTD component, we consider whether the star-formation timescales or mass-weighted ages of the sample affect the inferred DTD -- which would not be unexpected given the log-likelihood framework.
We discuss this result in more detail in the next section.
Finally, for case C the recovered DTD is as expected, i.e., a significant component in the final DTD bin ($>$10 Gyr) with weights in other bins close to zero. 

\begin{figure}
    \centering
    \includegraphics[width=\linewidth]{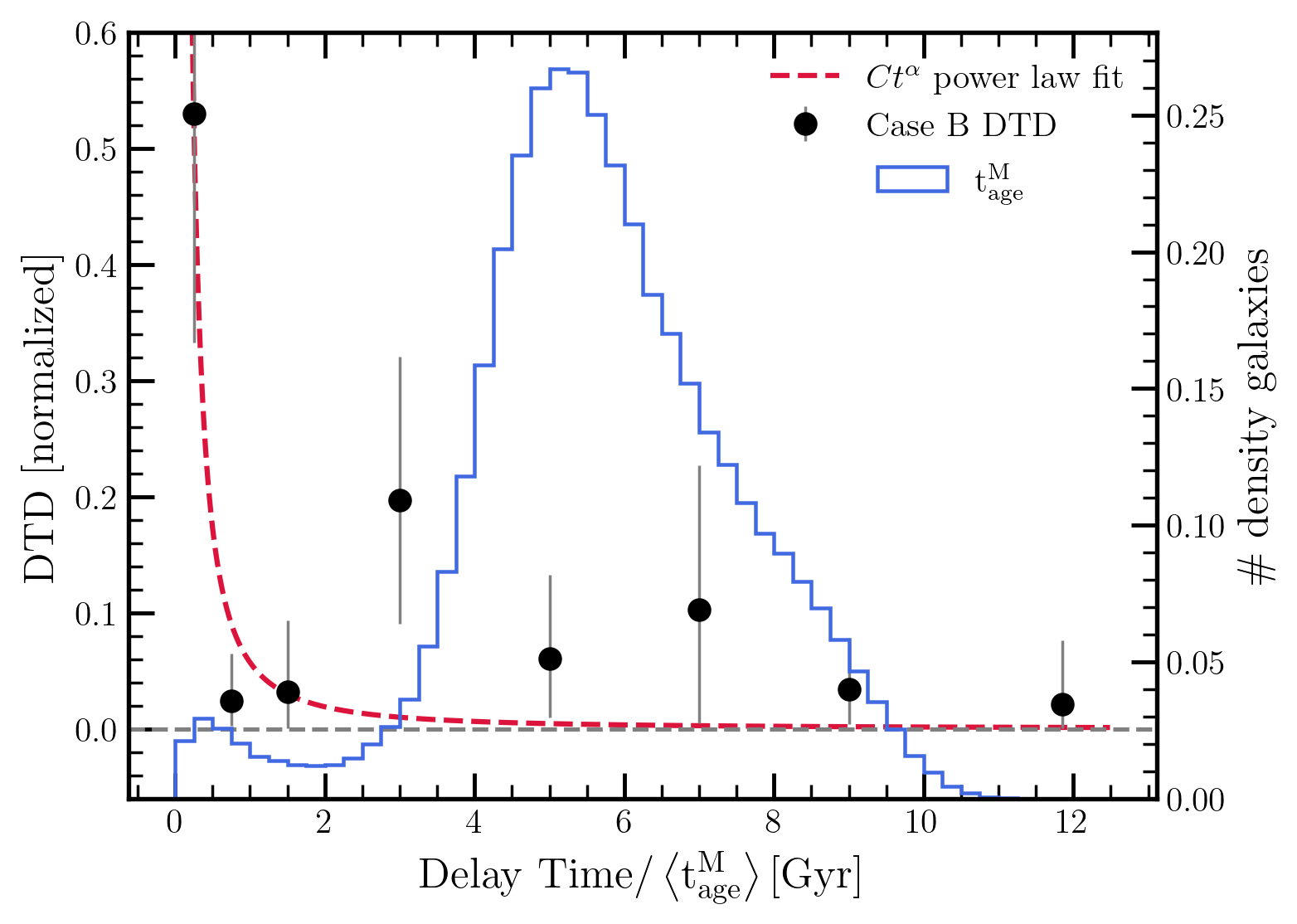}
    \caption{The DTD recovered for case B -- hosts with recent star formation (black points) -- shown along with the histogram of mass-weighted ages for the full sample. Also shown is a $Ct^\alpha$ power law fit to the DTD with best fit $\alpha = -1.6 \pm 0.68, C = 0.06 \pm 0.004$.}
    \label{fig:backage}
\end{figure}

\section{Discussion}
\label{sec:discussion}
The inferred DTDs in our test case B for late-time star-forming host galaxies, \emph{unexpectedly}, show significant weight within bins outside of the expected late-time bins (bins at $<$2 Gyr). While these hosts are chosen to be recently star-forming, however, there are no galaxies within our IllustrisTNG sample that have exactly zero star-formation beyond a lookback time of $\sim$2 Gyr (analogous to the case C hosts). Although these host galaxies show significant recent star formation they have also had star formation at lookback times $>$2 Gyr, as shown in the inset panel of case B in \cref{fig:dtd_rec_fullsample}.

If we were working with a real sample, these secondary components could potentially be hinting at multiple progenitor scenarios for SN~Ia \citep[e.g.,][]{Maoz14} with a dominant channel operating alongside other minor channels. However, the recovered DTDs in this work come from an artificial and extremely constrained selection of host galaxies. Therefore, it would be prudent to avoid physical interpretation through the DTDs recovered here. On the other hand, if the secondary components are a consequence of the maximum-likelihood exploration method we employ then this systematic feature could pose challenges when interpreting DTDs (from a real sample of galaxies) that are recovered with this method. 

Regardless of the cause, the shape of our recovered DTDs shows a more complex form than can be modeled with simple parametric expressions like a power law or an exponential. The assumption of a parametric form for the DTD would have difficulty uncovering the true DTD shape if it had \emph{both} prompt and delayed components with significant weights. We show that the commonly employed parametric form of a power law is a poor fit to the binned DTD in all our test cases in \cref{tab:summary}. The power law for the best fit in each case has significant uncertainties associated with it.
\cref{fig:backage} shows the best-fit power-law to the binned DTD from case B in more detail along with a histogram of the mass-weighted ages of the full 250K galaxy sample. It can also be seen that the secondary delay component at longer times (at $\sim$2--8 Gyr) shows broad agreement with the mass-weighted ages of the background population of galaxies. Below, we consider how such an entanglement between galaxy ages and delay times could come about.

With the DTD recovery method we are testing here, the DTD is determined through a maximum-likelihood exploration of the DTD parameter space. The convolution of the galaxy SFH and DTD (along with other multiplicative factors) determines the expected SN rate at the observed epoch for any given galaxy. We also know which observed galaxies hosted SN~Ia implying a peak in the SN~Ia rate history of these galaxies at the observed epoch. While exploring the parameter space DTD parameters that provide higher SN rates, at the observed epoch, matching the observed data are deemed likelier and up-weighted. Whereas other parameter values not matching observed data are deemed unlikely and down-weighted. However, we think the age bias arises because the average age at which individual galaxies formed their stellar mass is imprinted on the galaxy SFH which gets convolved with the DTD in the log-likelihood computation.
The mass-weighted age for any galaxy is a measure of the integrated SFH. It is defined for this work as:

\begin{equation}
    \mathrm{mass\ weighted\ age\, \equiv\, \left< t^M_{age} \right>} = \mathrm{t^{cosmos}_{age} (\emph{z})} - \frac{\int t\, \mathrm{SFH}(t) dt}{\int \mathrm{SFH}(t) dt},
\end{equation}

where $\mathrm{t^{cosmos}_{age} (\emph{z})}$ is the age of the Universe at redshift $z$, giving the mass-weighted age as a lookback time measure to enable better comparison to delay times. We suspect this causes the ages/average star-formation timescales to be imprinted on the DTD. Therefore, caution is advisable before interpreting the DTD inferred with this method from the MCMC exploration of log-likelihood constructed through individual galaxy SFHs.

\subsection{Limitations of this analysis}
While we have considered several common DTD cases, these cases, of course, do not constitute a complete sampling of the wide gamut of possibilities. For example, galaxies with multiple bursts/episodes of star-formation or those with star-formation affected by interactions/mergers are not considered. This is because the implications on the expected DTD outcome would be unclear for such cases.

There are also limitations of the IllustrisTNG simulation that are relevant to this work\footnote{See \url{https://www.tng-project.org/data/docs/background/\#sec3} for a more complete description of the limitations of IllustrisTNG.}.
Given the large scale of the cosmological simulations and the resolution limits, the IllustrisTNG simulation tracks star particles -- which are collections of stars that are all born at the same time with stellar masses distributed according to a Chabrier IMF \citep{Chabrier03}, i.e., a simple stellar population (SSP) \citep{Vogelsberger13, Pillepich18b}. The mass and metal return from a stellar particle are tracked with each timestep in the simulation. The predominant limitation for our analysis, given the resolution limits and the lack of an observational connection between SN~Ia and their progenitors, is the uncertainty on SN Ia rates. We do not know from observations what the SN Ia rate dependence is on stellar binary fractions, IMF, local metallicity, etc. Therefore, a calculation of SN Ia rates cannot be done within IllustrisTNG from first principles. As a result, IllustrisTNG uses a parameterization for the SN Ia rate based on the DTD prescription by \citet{Maoz12}. At each time step the mass and metal return from each stellar particle is computed according to the prescriptions for SN Ia, SN II, and AGB stars and returned to nearby cells.

This necessitates our choice of assigning host galaxies, as opposed to working with the hosts that IllustrisTNG might have chosen, by imposing requirements on their SFHs. The ``effective selection function" of supernova host galaxies in the real universe is likely a consequence of the interplay between many variables (e.g., local/global environment, host stellar mass, and perhaps redshift) -- which are all folded into the inferred DTD \citep[e.g.,][]{Graur17}.

\subsection{SN~Ia DTD from upcoming surveys}
The Vera C.~Rubin Observatory's Legacy Survey of Space and Time (LSST) is expected to discover millions of supernovae over its decade of operations \citep{Ivezic19}. The Nancy Grace Roman Space Telescope (Roman) High Latitude Time Domain Survey (HLTDS) is expected to find ${\sim}10^4$ SN~Ia with a significant fraction at $z{>}1$ \citep{Hounsell18, Akeson19}. Both of these surveys, along with data from the Euclid mission \citep{Laureijs11}, will significantly improve our understanding of dark energy. The LSST is expected to observe 18,000 deg$^2$ of the southern sky every few nights in the \emph{ugrizy} bands for 10 yr. The baseline strategy for LSST is expected to obtain light curves in six bands (\emph{ugrizy}) and photometric redshifts for ${\sim}400,000$ SN~Ia mostly at $z{<}1$ \citep{Ivezic19}, whereas the Roman HLTDS is expected to cover a few 10s of deg$^2$ with near-infrared photometry and slitless spectroscopy over a smaller fraction of the total survey area \citep{Rose21, Joshi22}.

This complementarity between Rubin LSST and Roman HLTDS should result in a large sample of SN~Ia with a broad redshift distribution. This sample will allow us to place meaningful constraints on the DTD over a wide range of redshift, environments, and host galaxy properties. In particular, we will be able to thoroughly test whether the SN~Ia DTD should have a dominant prompt component (as suggested by prior power law fits) or if it has significant components at longer delay times as well.

\begin{deluxetable*}{l c c c c c c c c c c}
\tablecaption{Summary of DTD results \label{tab:summary}}
\tabletypesize{\footnotesize}
\tablehead{\colhead{} & \multicolumn{8}{c}{DTD relative weights} &
\colhead{Best fit $C t^\alpha$} \\
\colhead{} & \colhead{Bin 1} & \colhead{Bin 2} & \colhead{Bin 3}
& \colhead{Bin 4} & \colhead{Bin 5} & \colhead{Bin 6} & \colhead{Bin 7} & \colhead{Bin 8}
& \colhead{power law [$C,\,\alpha$]}
}
\startdata
Case A1 &  0.013 $\substack{+ 0.01 \\ - 0.01 }$ & 0.002 $\substack{+ 0.01 \\ - 0.00 }$ & 0.014 $\substack{+ 0.01 \\ - 0.01 }$ & 0.057 $\substack{+ 0.04 \\ - 0.04 }$ & 0.046 $\substack{+ 0.03 \\ - 0.03 }$ & 0.355 $\substack{+ 0.16 \\ - 0.24 }$ & 0.403 $\substack{+ 0.12 \\ - 0.27 }$ & 0.110 $\substack{+ 0.07 \\ - 0.08 }$ & 0.04 $\pm$ 0.003, +0.76 $\pm$ 0.36 \\
Case A2 & 0.234 $\substack{+ 0.12 \\ - 0.08 }$ & 0.034 $\substack{+ 0.03 \\ - 0.02 }$ & 0.616 $\substack{+ 0.18 \\ - 0.14 }$ & 0.060 $\substack{+ 0.11 \\ - 0.04 }$ & 0.011 $\substack{+ 0.02 \\ - 0.01 }$ & 0.016 $\substack{+ 0.02 \\ - 0.01 }$ & 0.020 $\substack{+ 0.01 \\ - 0.01 }$ & 0.010 $\substack{+ 0.02 \\ - 0.01 }$ & 0.17 $\pm$ 0.008, -0.35 $\pm$ 0.17 \\
Case B & 0.529 $\substack{+ 0.19 \\ - 0.20 }$ & 0.024 $\substack{+ 0.04 \\ - 0.02 }$ & 0.032 $\substack{+ 0.06 \\ - 0.03 }$ & 0.197 $\substack{+ 0.12 \\ - 0.11 }$ & 0.060 $\substack{+ 0.07 \\ - 0.05 }$ & 0.103 $\substack{+ 0.12 \\ - 0.10 }$ & 0.034 $\substack{+ 0.04 \\ - 0.03 }$ & 0.021 $\substack{+ 0.06 \\ - 0.02 }$ & 0.06 $\pm$ 0.004, -1.60 $\pm$ 0.68 \\
Case C & 0.054 $\substack{+ 0.07 \\ - 0.04 }$ & 0.061 $\substack{+ 0.11 \\ - 0.05 }$ & 0.082 $\substack{+ 0.16 \\ - 0.07 }$ & 0.090 $\substack{+ 0.16 \\ - 0.07 }$ & 0.094 $\substack{+ 0.18 \\ - 0.07 }$ & 0.045 $\substack{+ 0.09 \\ - 0.04 }$ & 0.082 $\substack{+ 0.11 \\ - 0.05 }$ & 0.491 $\substack{+ 0.35 \\ - 0.31 }$ & $3.8 {\times} 10^{-7} {\pm} 4.8 {\times} 10^{-12}$, +5.69 $\pm$ 5.53\\
\enddata
\tablecomments{The delay time ranges, in Gyr, covered by the DTD bins are: 0--0.5, 0.5--1, 1--2, 2--4, 4--6, 6--8, 8--10, and 10--13.7 for bins 1 through 8, respectively.}
\end{deluxetable*}

\section{Summary}
\label{sec:summary}

This paper has attempted to address the following question: what influences, either from the data or the method, affect the inference of the SN~Ia DTD? There are many elements that contribute to answer the main question we present; the principal ones we have found include: the assumed form of the DTD (either parametric such as a power-law or non-parametric as the one used in this work) and the differences between the host galaxy SFHs vs the SFH for the average \emph{non-host} galaxy in the background population. Additional secondary effects, such as the number of DTD bins chosen and the delay time range covered by them, can also present themselves as systematic effects.

While \sne progenitor scenarios and delay times can successfully reproduce the observed SN rates, this work suggests that there may be different physical processes or evolutionary pathways leading to supernova explosions with different delay times, which should be accounted for in theoretical models and simulations. Continued efforts in observational studies, theoretical models, stellar population synthesis, and simulations are essential to address these challenges and improve our understanding of \sne progenitors.

Both cases B and C provide an important test for DTD recovery when the host galaxies are selected to have specific constraints on their SFHs. The tight constraints on host galaxy SFHs allow for a controlled framework of host vs \emph{non-host} galaxies and provide straightforward DTDs to expect before the run. We have shown that the form of the inferred DTD is driven in most part by the properties of the host galaxy SFH. We have also shown from Table~\ref{tab:summary} and Figure~\ref{fig:backage} that a power law is a poor fit in most cases.

One important future aspect is increasing the temporal resolution of the reconstructed DTD which will require interdisciplinary efforts spanning both observational and theoretical modeling. Advances in spectroscopic tools and data analysis methods are essential to overcoming these challenges and advancing our understanding of the parent systems and evolutionary trajectories \sne and their rates.
Therefore, this method of recovering the DTD and the potential systematic effects deserve a closer look, especially because larger samples of SN~Ia are expected in upcoming years from the Nancy Grace Roman Space Telescope and the LSST survey of the Rubin observatory.

\section *{Acknowledgements}
\begin{acknowledgments} \label{Sec:ack}

This work was supported by NSF-AAG award number 2205635.
This research has also made use of NASA’s Astrophysics Data System. YZ would like to thank Colin Norman for helpful discussions
\end{acknowledgments}
 
\software{astropy \citep{astropy}, Matplotlib \citep{matplotlib}, Numpy \citep{numpy}, and Scipy \citep{scipy}.}

\bibliographystyle{aasjournal}
\bibliography{references}

\appendix

\begin{figure}
    \centering
    \includegraphics[width=\textwidth]{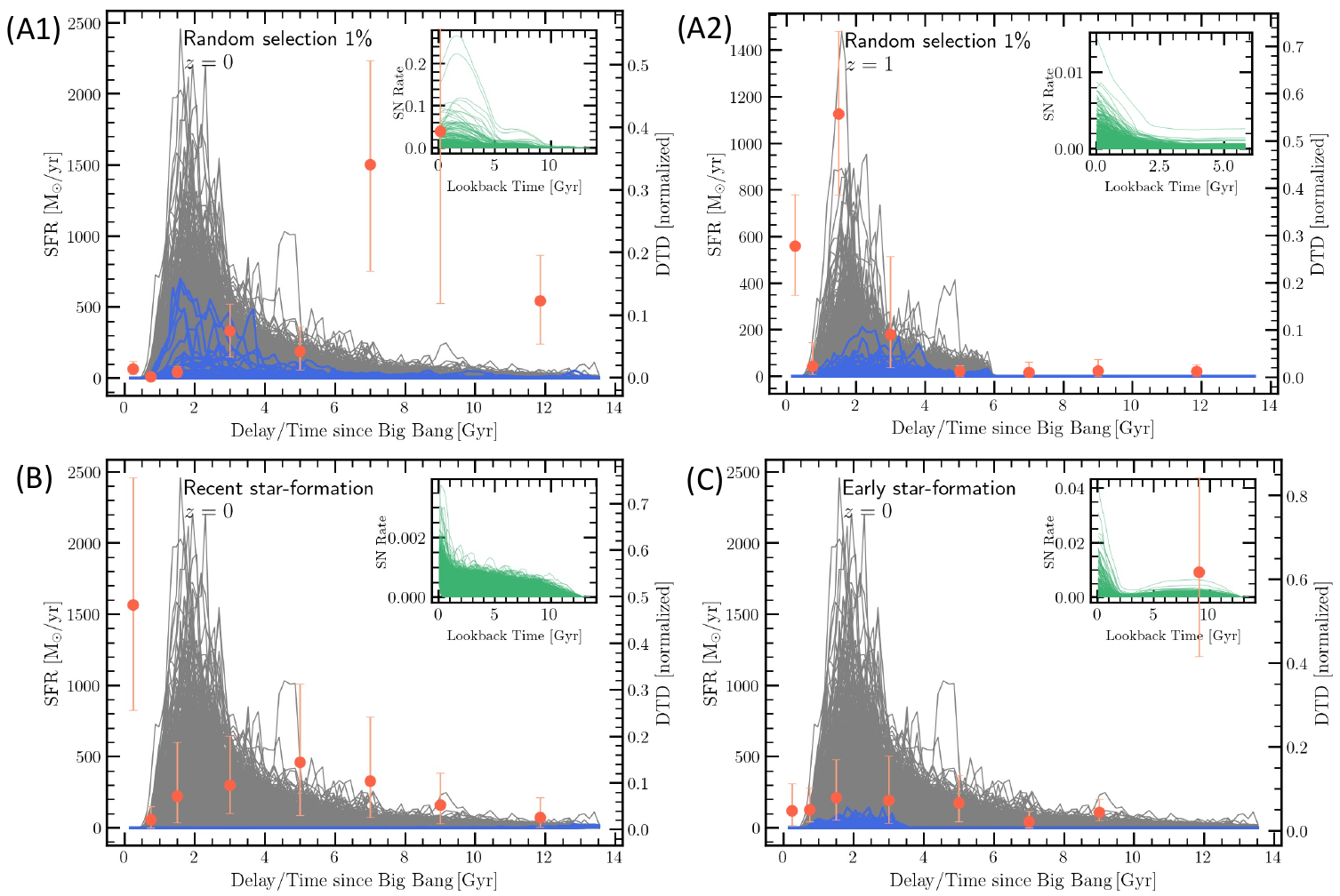}
    \caption{Same as \cref{fig:dtd_rec_fullsample} but with DTDs recovered by using a smaller fraction (20\%) of non host galaxies.}
    \label{fig:dtd_rec_reducedsample}
\end{figure}

Given the significant difference between the non host lnL term and the host log term (typically two to three orders of magnitude in our tests) we investigate if reliable DTD recovery is possible with a large reduction in the number of non host galaxies. This should allow for a significant reduction in computational expense without affecting the recovered DTD.
We test this idea by re-running all test cases on a reduced sample of galaxies. The sample is reduced such that all original hosts are kept but the \emph{non-host} galaxies are reduced to 20\% of their initial number.
The DTD results from the reduced sample are shown in \cref{fig:dtd_rec_reducedsample}.

The significant reduction in the computational load comes from replacing the term for \emph{all} \emph{non-hosts} with a term for a \emph{representative sample of non-host} galaxies. By reducing the number of non-hosts we essentially reduce two computations that come with significant cost: (i) computing SFHs for non-host galaxies and (ii) calculating SN rates for non-host galaxies in the lnL term for each DTD parameter set tested which must be done iteratively as each MCMC walker computes the lnL for each proposed parameter set.

\end{document}